\documentclass[conference]{IEEEtran}
\ifCLASSINFOpdf
\else
\fi
 \usepackage{graphicx}

\begin{document}
%
\title{3D/4D Ultrasound Registration of Bone}




%
\author{\IEEEauthorblockN{Jonathan Schers\IEEEauthorrefmark{1}\IEEEauthorrefmark{3},
Jocelyne Troccaz\IEEEauthorrefmark{1},
Vincent Daanen\IEEEauthorrefmark{2},
C\'eline Fouard\IEEEauthorrefmark{1}, 
Christopher Plaskos\IEEEauthorrefmark{3} and
Pascal Kilian\IEEEauthorrefmark{3}}
\IEEEauthorblockA{\IEEEauthorrefmark{1}TIMC-IMAG Laboratory, GMCAO Department\\
La Tronche, FRANCE\\ Email: Jonathan.Schers@imag.fr}
\IEEEauthorblockA{\IEEEauthorrefmark{2}Formerly TIMC-IMAG Laboratory, GMCAO Department\\
La Tronche, FRANCE}
\IEEEauthorblockA{\IEEEauthorrefmark{3}Praxim-Medivision\\
La Tronche, FRANCE}}


\maketitle

\begin{abstract}
This paper presents a method to reduce the invasiveness of Computer Assisted Orthopaedic Surgery (CAOS) using ultrasound. In this goal, we need to develop a method for 3D/4D ultrasound registration. The premilinary results of this study suggest that the development of a robust and ``realtime'' 3D/4D ultrasound registration is feasible.
\end{abstract}


%
\IEEEpeerreviewmaketitle

\section{Introduction}

\begin{figure}[!t]
	\centering \includegraphics[width=3.5in]{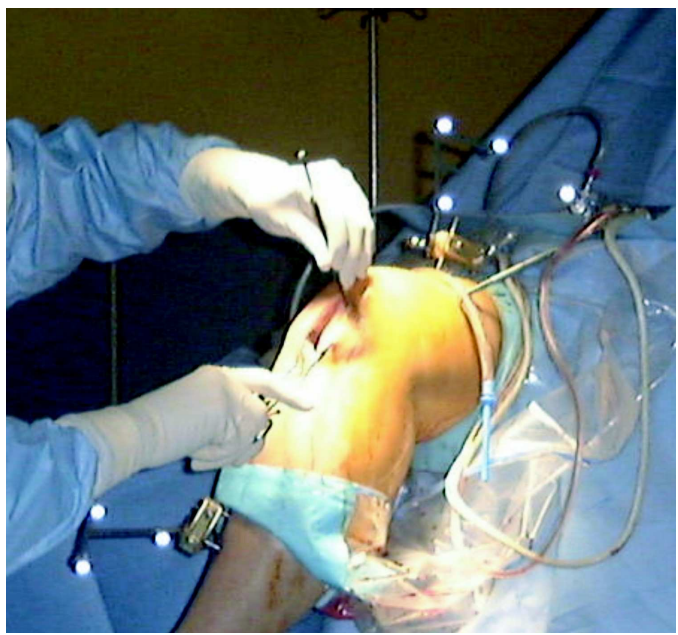}%
	\caption{Optical trackers which are rigidly pinned in to the bone during an anterior cruciate ligament surgery.}
	\label{fig_RB}
\end{figure}

The Development of 3D ultrasound offers interesting prospects for surgical navigation. Indeed, this non-invasive data and ``realtime'' imaging technology could reduce the invasiveness of Computer Assisted Orthopaedic Surgery (CAOS) by replacing the optical trackers which, when pinned into the bones (Fig. \ref{fig_RB}), could increase the risk of infection and cause extra pain owing to extra incisions and exposed bone surfaces. This implies the development of a method for rigid registration in order to track a bony structure. This method has to be at the same time robust and ``realtime''.

However, due to the speckles, the shadowing effects and the poor quality of the images, the registration of this imaging modality is a challenging process. Consequently, the research on intra-modality registration of ultrasound images is scarcely reported in the literature \cite{CARS-LetteboerVN03}, \cite{IEEEMI-ShekharZ02}, but there appears to be a greater interest in work related to the application of ultrasound images \cite{DBLP:journals/tmi/PluimF03}.

To our Knowledge there is no report on the evaluation of a method for tracking bone structures whitout invasive trackers.

\begin{figure}[!t]
	\centering \includegraphics[width=3.5in]{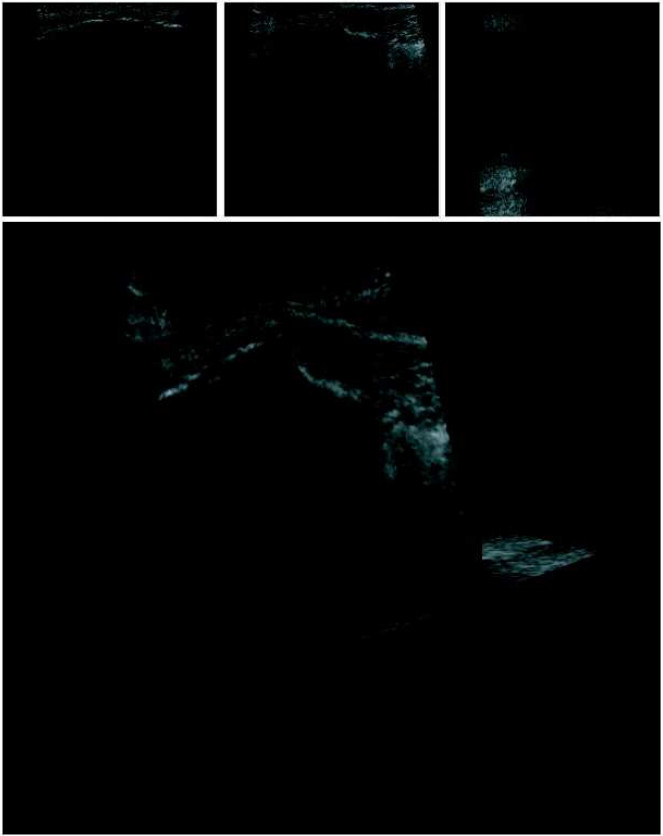}%
	\caption{4D ultrasound image of an anterior tibial tuberosity, i.e. 3 orthogonal planes obtained in ``realtime''.}
	\label{fig_Brutes}
\end{figure}

This article is organized as follows. Section 2 presents the material and the method. Evaluation and results are presented in section 3. Section 4 concludes the article and deals with the future work.

\section{Material and Method}

\subsection{Material}
Ultrasound data have been obtained with a General Electric Voluson 730 Pro, equipped with a 3D RSP 6-16 MHz probe. The size of the ultrasound volumes is $199 \times 199 \times 199$ voxels, and the voxel size is 0.28 mm. Each volume acquisition take almost 5s.

\subsection{Method}
The essential idea of the method that we propose consists in producing a coarse segmentation of the interface between the bone and the soft tissues to select a region of interest around the structures containing information.

\subsubsection{Automatical coarse segmentation}
\begin{figure}[!t]
	\centering \includegraphics[width=3.5in]{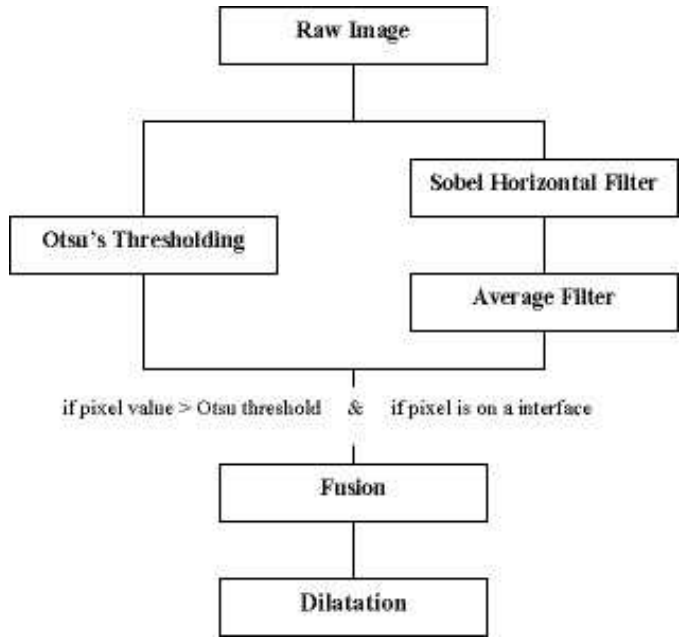}%
	\caption{Image processing to obtain the automatical coarse segmentation.}
	\label{fig_Diagramme}
\end{figure}

Several types of information, based on the physics of the ultrasound imaging -- (1) the bone appears to be hyper-echoic because the difference between the acoutic impedance of the bone and the soft tissues is important, (2) due to the bones have a high absorption rate and there is no imaging possible beyong them, an acoustic shadow is found behind the interface and (3) only the more or less horizontal interfaces can be seen in images because the reflexion of ultrasound is almost specular --, have been used in four different steps. The differents steps have been summarized in the Fig. \ref{fig_Diagramme}.

In a first step, the images have been binarized according to Otsu's thresholding method which allows a good approximation of the separation between the echogenic zone and the shadow zone \cite{DBLP:conf/eccv/DaanenTT04}.

In a second step, a horizontal sobel filter has been applied to detect interfaces. We have decided to use this filter because the bone interfaces which can be found are nearly horizontal.

In a third step, the images have been averaged to remove the noise. And, eventually, fusioning has been used relying on two conditions: 

\begin{itemize}
\item the pixel value is greater than Otsu's threshold;
\item the pixel belongs to an interface (i.e. it is a local maximum on the ultrasound beam).
\end{itemize}

A dilatation has also been performed to increase the number of selected pixels. The results of this processing is a region of interest containing information.

\subsubsection{Registration}

The performance of using a voxel-based registration for image registration has been recognized in the literrature \cite{maintz98survey}.
Consequentely, and  because to obtain an accurate segmentation of the bone in ultrasound imaging for a surface-based registration is a challenging process, a voxel-based registration approach has been used.

The images have then been compared according to the normalised cross-correlation (NCC) similarity measure. We have chosen this similarity measure because in the literature \cite{MP-KimFLBT01}, this measure seems to be the best for monomodality registration. $\bar{A}$ represents the mean value of the image $A$ and $\bar{B}$ represents the mean value of the image $B$. The NCC is defined by as follow:

$$ NCC = \frac{\displaystyle \sum_{i} \left ( A\left( i \right ) - \overline{A} \right ) \left ( B'\left ( i \right ) - \overline{B'} \right ) }{\sqrt{\displaystyle \sum_{i} \left ( A\left ( i \right ) - \overline{A}\right ) ^{2} \sum_{i} \left ( B'\left ( i \right ) - \overline{B'} \right )^{2} }} $$

Among the optimization algorithms, we have selected the simplex method of Nelder and Mead \cite{simplex} for its robustness and convergence time that's why this method is often used for image registration \cite{UMB-SlomkaMDF01}, \cite{IEEEMI-ShekharZ02}, \cite{MP-RadauSJSW01}, \cite{MIA-MaesVS99}, \cite{UMB-MeyerBKBLFRC99} and \cite{MP-KimFLBT01}. This method optimizes all the parameters in the same time. The simplex method of Nelder and Mead consists in updating a $n$-dimensional simplex such as the value of the similarity measure decrease. At each step, the simplex is modified by geometric operations. A new vertex is accepted according to the value of the similarity measure.

The size of the initial simplex in a parameter space is an important step using simplex optimization. For rigid transformation the space is 6-dimensional. Before determining the size of the initial simplex, we need to process a normalization so that the displacement in the parameters space is approximately the same as in the spatial space. This relationship is checked for translation. However, it is not true for rotation, where the displacement of a voxel is dependent on its distance from the axis rotation. In the data sets we have used in this study, a unit parameter corresponded to 1 mm of translation and 1$^\circ$ of rotation.

There are two conditions for the size of the initial simplex:
\begin{itemize}
	\item In principle, it should be greater than the unit dimension along each axis so that it does not get stuck in a local minimum;
	\item It should not be greater than the capture range, otherwise, the convergence may not occur.
\end{itemize}

The size in each direction has been initialized between 3 and 5 units.

The initial attitude corresponds to the result of the previous optimization (for each sequence of images) if registration is successful.

\section{Evaluation and Results}

\subsection{Evaluation method}

\begin{figure}[!t]
	\centering \includegraphics[width=3.5in]{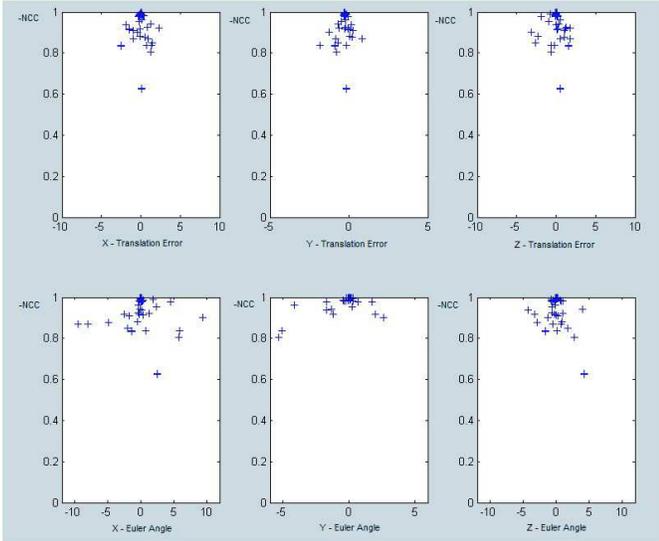}
	\caption{Correlation beetween $T_{error}$ and the similarity measure found during registration (60 images).}
	\label{fig_Graphe}
\end{figure}

In order to evaluate our method, we simulated 4D data by reslicing orthogonal 2D slices from the 3D ultrasound volumes. This simulated data were translated and rotated in the maximum range of $\pm$10mm --except according to the probe's axis $\pm5mm$-- / $\pm$12$^\circ$ --except according to the perpendicular axis of the probe's axis $\pm$6$^\circ$-- which represents a known reference transformation $T_{reference}$. The registration of the 4D data to the reference volume provides a transformation $T_{registration}$. The composition of $T_{reference}$ and $T_{registration}$ gives a transformation error $T_{error}$ of the $T_{registration}$:
$$ T_{error} = T_{reference}^{-1}T_{registration}$$

From this transformation error $T_{error}$ we extract a translation vector and 3 Euler angles which correspond to the error in each direction.
It should be noted that $T_{error}$ is correlated to the similarity measure $\alpha$ found during registration (Fig. \ref{fig_Graphe}):  when the error is less than 1mm and 1$^\circ$ then $\alpha$ is closed to the absolute maximum, i.e. $0.95\leq \alpha \leq 1$.

So as to assess our results, we have defined 2 criteria allowing to validate a registration:
\begin{itemize}
\item the similarity measure is over 0.95;
\item the errors  in translation and rotation are inferior to 1 mm and 1 $^\circ$.
\end{itemize}

In the following sections, we will maintain that a registation is proved successful if these 2 critera are estabilished.

\subsection{Results}

The proposed method has been tested on ultrasound knee images. Three data sets have been used. Each set contains 60 images (i.e. 2 orthogonal slices). The table \ref{table_example} shows the results which we have obtained. In our data, the registration that we propose is successful in almost 65\% of the cases, what is encouraging for the future work. Furthermore, the average time per registration is almost 10s per images, what lets imagine than ``realtime'' is possible. Indeed, our code is not optimized.

\begin{table}[!t]
\renewcommand{\arraystretch}{2.0}
\caption{Percentage of success for each data sets (60 images per data sets)}
\label{table_example}
\centering
\begin{tabular}{|c||c|c|}
\hline
Data Sets & Success (\%) & Total Time (s)\\
\hline\hline
1 & 68.3 & 696\\
\hline
2 & 63.3 & 703\\
\hline
3 & 61.7 & 729\\
\hline
\end{tabular}
\end{table}

\section{Conclusion and Future work}
This first evaluation of the method shows that rigid registration of 3D/4D ultrasound data works in approximately 65\% of the cases in a reasonable time. The average time of a registration is approxymately 10s. These initial results suggest that the development of a robust and ``realtime'' 3D/4D ultrasound registration method is feasible. 

Current work deals with improvment of accuracy and the robustness.

We also need to evaluate and compare our method with a ``gold standard''. The ``gold standard'' will be defined by the optical trackers pinned in to the bones.






%

\IEEEtriggeratref{10}
\bibliographystyle{IEEEtran}
\bibliography{Article}

\end{document}